\pdfoutput=1
\documentclass[11pt]{article}
\usepackage[english]{babel}
\usepackage{graphicx,amsfonts,amsmath,amssymb,mathrsfs}
\usepackage[caption=false]{subfig}
\usepackage[titletoc,title]{appendix}
\usepackage{a4wide}
\usepackage[colorlinks=true
,urlcolor=blue
,anchorcolor=blue
,citecolor=blue
,filecolor=blue
,linkcolor=blue
,menucolor=blue
,linktocpage=true]{hyperref}

\def\p{\partial}

\def\sign{\qopname\relax o{sign}}
\def\Vol{\qopname\relax o{Vol}}

\renewcommand{\i}{\mathrm{i}}
\renewcommand{\d}{\mathrm{d}}

\newcommand{\beq}{\begin{eqnarray}}
\newcommand{\eeq}{\end{eqnarray}}
\newcommand{\non}{\nonumber\\}

\newcommand{\bOmega}{\boldsymbol{\Omega}}
\newcommand{\zb}{\bar{z}}
\newcommand{\eb}{\bar{e}}
\newcommand{\type}[3]{type #1$_{\texttt{#2}}^{\texttt{#3}}$}
\newcommand{\Type}[3]{Type #1$_{\texttt{#2}}^{\texttt{#3}}$}

\begin{document}
\title{Nineteen vortex equations and integrability}
\author{Sven Bjarke Gudnason\\[16pt]
  {\small Institute of Contemporary Mathematics,}\\
  {\small School of Mathematics and Statistics,}\\
  {\small Henan University, Kaifeng, Henan 475004, P.~R.~China}
}
\date{July 2022}
\maketitle

\begin{abstract}
  The class of five integrable vortex equations discussed
  recently by Manton is extended so it includes the relativistic BPS
  Chern-Simons vortices, yielding a total of nineteen vortex equations.
  Not all the nineteen vortex equations are integrable, but four new
  integrable equations are discovered and we generalize them to
  infinitely many sets of four integrable vortex equations, with each
  set denoted by its integer order $n$.
  Their integrability is similar to the known cases, but give
  rise to different (generalized) Baptista geometries, where the
  Baptista metric is a conformal rescaling of the background metric by
  the Higgs field.
  In particular, the Baptista manifolds have conical singularities.
  Where the Jackiw-Pi, Taubes, Popov and Ambj{\o}rn-Olesen vortices
  have conical deficits of $2\pi$ at each vortex zero in their
  Baptista manifolds, the higher-order generalizations of these
  equations are also integrable with larger constant curvatures and a
  $2\pi n$ conical deficit at each vortex zero.
  We then generalize a superposition law, known for Taubes vortices of
  how to add vortices to a known solution, to all the integrable
  vortex equations. We find that although the Taubes and the Popov
  equations relate to themselves, the Ambj{\o}rn-Olesen and Jackiw-Pi
  vortices are added by using the Baptista metric and the Popov
  equation.
  Finally, we find many further relations between vortex equations,
  e.g.~we find that the Chern-Simons vortices can be interpreted as
  Taubes vortices on the Baptista manifold of their own solution.
\end{abstract}
\indent\texttt{email:gudnason@henu.edu.cn}\\

\vspace{2pc}
\noindent{\it Keywords}: Abelian vortex equations, Baptista metric, conical
singularities, Liouville equation

\newpage
\tableofcontents

\section{Introduction}

Abelian vortices play an important role as perhaps the simplest
nontrivial solitons in gauge theories and have applications in a
variety of fields, e.g.~in superconductors or as cosmic strings.
The relativistic generalization of Ginzburg-Landau theory describing
the vortices in type II superconductors is called the Abelian Higgs
model.
It has a single parameter describing the physics of the model, namely
the ratio of the Higgs mass to the photon mass.
When the ratio is unity, the vortices are said to be critically
coupled and their governing equations reduce from two coupled
second-order partial differential equations (PDEs) to two first-order
PDEs; the latter two can then be combined, yielding a single vortex
equation of second order, which is known as the Taubes equation
\cite{Taubes:1979tm,JT,MS,YangsBook}.

The Taubes vortex equation, although simpler than the generic coupling
case, is not integrable and hence no closed-form analytic expression
for its solution is known.
Modification of the base geometry, however, changes this fact
and as first found by Witten, the Taubes equation becomes integrable
when the base manifold is tuned to a particular constant
negative curvature, turning the manifold into a hyperbolic plane
\cite{Witten:1976ck}.
Integrability requires that the curvature of the base manifold
matches the constant term in the Taubes equation.
Further integrable vortex equations are known, such as that describing
the Jackiw-Pi vortices in nonrelativistic Chern-Simons theory
\cite{Jackiw:1990tz} and that describing Popov vortices
\cite{Popov:2012av}.
These three vortex equations and further two integrable vortex
equations for the Ambj{\o}rn-Olesen \cite{Ambjorn:1988fx} and Bradlow
vortices, were then put on equal footing by Manton
\cite{Manton:2016waw}.

The five vortex equations studied by Manton can nicely be understood
geometrically in terms of a quantity called the Baptista metric, which
is the background metric after a conformal rescaling by the Higgs
field squared, yielding $g_2=g_0|\phi|^2$.
The integrability of the vortex equations are then understood as the
condition that the background Gauss curvature is constant and equal to
the constant term in the vortex equation and the curvature
corresponding to the Baptista metric is constant and equal to the
coefficient of the Higgs field squared. Up to rescaling of lengths and
field redefinitions, the types of integrable vortex equations are thus
simply given by the signs of these two coefficients -- in this paper
we shall call them vortex coefficients (VCs). 

Although it turns out that all the possible vortex equations that one
can write down satisfying the positive (magnetic) flux condition and
being linear in the Higgs field squared (i.e.~$|\phi|^2$) are
integrable, one may wonder why stop at the linear order in $|\phi|^2$?
In fact, a well known vortex equation (at critical coupling) that is
not linear but quadratic in $|\phi|^2$ is the relativistic BPS
Chern-Simons vortex equation \cite{Jackiw:1990aw}.
It is, however, not integrable.
Nevertheless, in this paper we will extend the vortex equations from
the five vortex equations of Manton to the next order in $|\phi|^2$,
yielding nineteen vortex equations and the title of the paper.
If we also include the vanishing vortex polynomial, which we shall
call the Laplace vortex equation, there will be twenty vortex
equations.

In this paper, we first perform a classification of the vortex
equations and count how many types of vortex equations are possible
under the condition that the magnetic flux remains positive definite
as a function of the highest power of the Higgs field squared, $L$.
The nineteen or rather twenty vortex equations studied in this paper
correspond to $L=2$ and includes the Chern-Simons equation.
One may expect that the equations become increasingly more difficult
and less likely to be integrable, but we find that for each increment
of $L$ by one, there are 4 new integrable vortex equations; in
particular they can be thought of as the higher-order generalizations
of the Jackiw-Pi, the Taubes, the Popov and the Ambj{\o}rn-Olesen
vortex equations.
They can be understood geometrically by generalizing the Baptista
metric to a higher-order Baptista metric as $g_{2n}=g_0|\phi|^{2n}$.
The new integrable vortex equations correspond to cases where the
higher-order Baptista manifolds have constant curvature.
The higher order of the metric also impacts the geometry locally as
the higher-order Baptista manifolds for integrable vortices are
constantly curved spaces with conical singularities, where each single
vortex gives rise to a conical excess of $2\pi n$, with $n$
corresponding to the order of the Baptista manifold.
We also integrate the newly found integrable vortex equations, but the
corresponding Bradlow-type bounds turn out to be independent of the
order of the Baptista manifold.
We then consider superposition rules and the geometric interpretation
of adding vortices to known vortex solutions for all of the integrable
types of vortices, generalizing the known result of Baptista for
Taubes vortices.
Finally, we contemplate further relations among the vortex equations
and find for example that the Chern-Simons vortex can be interpreted
geometrically as a Taubes vortex on the Baptista background of
itself.
Using this relation we contemplate the geometry that the integrable
solutions of the Taubes equation gives rise to for the Chern-Simons
vortices, which turns out to be singular.

This paper is organized as follows.
In Sec.~\ref{sec:vtxeqs} we set up the geometry, the vortex equations
and the Baptista metric.
In Sec.~\ref{sec:types} we classify the vortex equations for $L=2$,
i.e.~the simplest twenty vortex equations on Taubes form.
In Sec.~\ref{sec:integrability} we review the five integrable vortex
equations of Ref.~\cite{Manton:2016waw} and present four new ones.
In Sec.~\ref{sec:volumes} we integrate the vortex equations and find
Bradlow-type bounds that for some vortex equations limit the number of
vortices for a given volume, but put a lower bound on others.
In Sec.~\ref{sec:deficits} we prove that the vortex equations imply an
analytic form of the solutions in a small open disc around a vortex
zero and use this result to calculate the conical deficits of the
Baptista manifolds at those vortex zeros.
In Sec.~\ref{sec:superposition} we generalize the vortex superposition
law of Baptista for Taubes vortices to all the integrable vortex
equations, including the other four equations of
Ref.~\cite{Manton:2016waw} as well as the newly found integrable
vortex equations.
A summary of our relations and results can be seen in
Fig.~\ref{fig:relations}.
In Sec.~\ref{sec:singular}, we find the singular geometry for which
the integrable Taubes vortices on the hyperbolic plane give rise to
Chern-Simons vortices. 
Finally, we conclude the paper with a discussion and outlook in
Sec.~\ref{sec:discussion}.

\section{Vortex equations on Riemann surfaces}\label{sec:vtxeqs}

\subsection{The geometry}

Consider a Riemann surface $(M_0,g_0)$, equipped with a metric $g_0$,
having a constant Gauss curvature $K_0$ and we will use a local
complex coordinate $z$.
The Riemann surface has the metric $(ds_0^2)$:
\beq
g_0 = \Omega_0 \d z\d\zb,
\eeq
with $\Omega_0$ being the conformal factor
\beq
\Omega_0 = \frac{4}{(1+K_0| z|^2)^2},
\eeq
admits a local complexified frame
\beq
e = \frac{2}{1+K_0| z|^2}\d z \in\bOmega^{1,0}(M_0),
\eeq
which obeys the structure equation
\beq
\d e - \i\Gamma\wedge e = 0,
\eeq
and $\Gamma$ is the spin connection 1-form
\beq
\Gamma = \i K_0\frac{z\d\zb - \zb\d z}{1+K_0| z|^2}\d\zb \in\bOmega^{1}(M_0),
\eeq
where we denote generic $r$-forms on $M_0$ by $\bOmega^r(M_0)$ and
holomorphic $r$-forms on $M_0$ by $\bOmega^{r,0}(M_0)$. 
The curvature 2-form is related to the spin connection via
\beq
R = \d\Gamma = \frac{\i}{2}K_0 e\wedge\eb \in\bOmega^{1,1}(M_0),
\eeq
with $K_0$ being the Gauss curvature, which can also be obtained
directly from the conformal factor of the metric $g_0$ on $M_0$ as
\beq
K_0 = \frac12\Delta_{g_0}\log\Omega_0
= \frac12(\d\delta+\delta\d)\log\Omega_0
= -\frac{2}{\Omega_0}\p_z\p_{\zb}\log\Omega_0,
\label{eq:K0}
\eeq
where $\delta=-*\d*$ is the coderivative and $*$ is the Hodge
``star'' operation, $*:\bOmega^{s,p}(M_0)\to\bOmega^{1-p,1-s}(M_0)$.
$\Delta_{g_0}=\d\delta + \delta\d$ is the Hodge Laplacian with respect
to the metric $g_0$. 
$\p_z=\frac{\p}{\p z}$ and $\p_{\zb}=\frac{\p}{\p\zb}$ are partial
derivatives of the local coordinates.

\subsection{Vortex equations}

We will denote vortices by the pair $(\phi,A)$ of a connection 1-form
$A$ and a smooth section $\phi$ on a line bundle over the Riemann
surface $M_0$.
The vortex equation will be taken to be
\begin{align}
  \d_A\phi\wedge e &= (\d\phi - \i A\phi)\wedge e = 0,\label{eq:BPS1}\\
  F &= \d A = P(|\phi|^2) \omega_0,\label{eq:BPS2}
\end{align}
where $\omega_0=\frac{\i}{2}\Omega_0\d z\wedge\d\zb\in\bOmega^{1,1}(M_0)$ is the K\"ahler form on
$M_0$ and $P(|\phi|^2)$ is the vortex polynomial.
Using the Maurer-Cartan solution of Eq.~\eqref{eq:BPS1},
\beq
A_{\zb} = -\i\p_{\zb}\log\phi,
\eeq
and the decomposition of the Higgs field $\phi=e^{u+\i\chi}$, we
arrive at the vortex equation on ``Taubes'' form
\beq
\Delta_{g_0} u = 
-\frac{4}{\Omega_0}\p_z\p_{\zb}u = P(e^{2u})
- \frac{2\pi}{\Omega_0}\sum_{i=1}^N\delta^{(2)}(z-z_i),
\eeq
where $z_i$, $i=1,\ldots N$ are zeros of the Higgs field $\phi$.
We will denote the set of not-necessarily distinct zeros by $\{z_i\}$
and the distinct zeros $z_k\in\{z_i\}$ with multiplicities $N_k$.
The sum over Dirac-delta functions will be abbreviated by
\beq
\delta_D := 2\pi\sum_{i=1}^N\delta^{(2)}(z - z_i),
\eeq
where $D=\phi^{-1}(0)\equiv\{z_i\}$, $i=1,\ldots,N$ is the effective divisor.
For reference, we note that the field strength 2-form is given by
\beq
F = -2\i\p_z\p_{\zb}u\,\d z\wedge\d\zb \in\bOmega^{1,1}(M_0),
\eeq
whereas the magnetic flux is
\beq
B = *F
= -\frac{4}{\Omega_0}\p_z\p_{\zb}u = \Delta_{g_0}u.
\eeq
The integral of the magnetic flux is the first Chern class of the line
bundle
\beq
\frac{1}{2\pi}\int_{M_0} F = N.
\label{eq:N}
\eeq
Restricting to $N>0$ positive, without loss of generality, we also
have that
\beq
0<N = -\frac{2}{\pi}\int\p_z\p_{\zb}u\;\d x^1\wedge\d x^2,
\eeq
which implies that
\beq
\int_{M_0}P(e^{2u})\omega_0 > 0,
\label{eq:positive_flux_condition}
\eeq
which we henceforth shall call the positive flux condition.

\subsection{The Baptista metric}

A geometric interpretation of the vortex moduli space is given by
considering the Baptista metric, which we will denote by $g_2$ and may
define as \cite{Chen:2004xu,Baptista:2012tx,Manton:2016waw}
\beq
g_2=\Omega_2\d z\d\zb,\qquad
\Omega_2 := \Omega_0|\phi|^2 = \Omega_0 e^{2u},
\eeq
which is, of course, degenerate at the zeros of the Higgs field
$\{z_i\}$.
Its corresponding curvature is given by
\begin{align}
K_2 &= \frac12\Delta_{g_2}\log\Omega_2
= -\frac{2}{\Omega_2}\p_z\p_{\zb}\log\Omega_2 \non
&= \frac{\Omega_0}{\Omega_2}\left(K_0 + \Delta_{g_0}u\right)
= \frac{\Omega_0}{\Omega_2}\left(K_0 - \frac{4}{\Omega_0}\p_z\p_{\zb} u\right),
\label{eq:K2}
\end{align}
which can be expressed in terms of the background curvature $K_0$ and
the magnetic flux 2-form $F$;
it is only defined on $M=M_0\backslash\{z_i\}$.
In particular, the metric possesses conical singularities at the
points $\{z_i\}$ where the Higgs field vanishes
\cite{Baptista:2012tx,Manton:2016waw}.

We will now define a generalized Baptista metric, which is given by
\beq
\Omega_{2n} \equiv \Omega_0 e^{2n u},
\eeq
where $n\in\mathbb{Z}_{\geq 0}$ is a positive semi-definite integer.
For $n>0$, it is degenerate at the points $\{z_i\}$. 
The trivial case, $n=0$, is just the background metric of course and
is regular.
For $n=1$ it is the normal Baptista metric and $n=2$ is the
highest-order Baptista metric that we will consider in this paper.
The corresponding Gauss curvatures then become
\begin{align}
K_{2n} &= \frac12\Delta_{g_{2n}}\log\Omega_{g_{2n}}
= -\frac{2}{\Omega_{2n}}\p_z\p_{\zb}\log\Omega_{2n} \non
&= \frac{\Omega_0}{\Omega_{2n}}\left(K_0 + n\Delta_{g_0}u\right)
= \frac{\Omega_0}{\Omega_{2n}}\left(K_0 - \frac{4n}{\Omega_0}\p_z\p_{\zb}u\right),
\label{eq:K2n}
\end{align}
and are only defined on $M=M_0\backslash\{z_i\}$, except for $n=0$.
Notice that there are nontrivial relations between the higher-order
Baptista metrics, e.g.,
\beq
\Omega_4 = \frac{\Omega_2^2}{\Omega_0}.
\eeq
An interesting relation can be found by combining the curvatures
$K_2$ of Eq.~\eqref{eq:K2} and $K_4$ of Eq.~\eqref{eq:K2n} with $n=2$
in such a way as to eliminate the field strength 2-form, yielding
\beq
\Omega_0 K_0 = 2\Omega_2 K_2 - \Omega_4 K_4.
\eeq
It illustrates what is clear from the construction, namely that the
curvatures of the Baptista manifolds are related to each other via the
background curvature.

\subsection{The vortex polynomial}

The vortex polynomial, $P$, for the standard Taubes vortices is
proportional to the square root of the Higgs potential.
In this paper, we will not concern us with the underlying theory
giving rise to the equations or whether they pose a well-defined
variational problem and if they do, in what is the well-defined
underlying theory that does.

The vortex polynomial may be written as
\beq
P(e^{2u}) = -C_0 + \sum_{n=1}^L C_{2n} e^{2n u},
\label{eq:Pseries}
\eeq
with $C_{2n}\in\mathbb{R}$ and $e^{2u}$ being real coefficients and
the square of the norm of the Higgs field.

$L=1$ corresponds to the case covered in Ref.~\cite{Manton:2016waw}
giving rise to $Q=5$ vortex equations.
If the case of $C_{2n}=0\;\forall n$ is included, there are $Q=6$ vortex
equations at $L=1$.
The vortex equation with a vanishing vortex polynomial, $P=0$, was
considered in Ref.~\cite{Contatto:2017alh} and the corresponding
vortices may be called Laplace vortices.

In this paper, we will take $L=2$ motivated by the Chern-Simons vortex
equation at critical coupling, which corresponds to $C_0=0$, $C_2=1$
and $C_4=-1$.
We will see shortly that this case gives rise to $Q=19$ vortex
equations as well as the title of the paper or rather $Q=20$ vortex
equations, if we include the Laplace vortices.
We can calculate the number of vortex equations $Q$ as a function of
the highest power, $L$ of $e^{2u}$, in the series \eqref{eq:Pseries}
as follows. 
The total number of possibilities is classified by each coefficient
being positive, negative or zero\footnote{When three or more
  coefficients are nonvanishing, there will be families of equations;
  however, we will still classify them according to the signs of the 
  coefficients that cannot be fixed in magnitude by rescaling. }.
Counting the number of possibilities is rather easy. The relevant
space is the sign function
\beq
\sign : \mathbb{R} \to \{-1,0,+1\}\equiv\oplus,
\eeq
where we have defined the latter symbol to be the set of $-1$, $0$ and
$+1$.
For each term in the polynomial, there are thus 3 possible values of
the coefficient (in our classification of the equations) and hence
$3^{L+1}$ different equations (recall that the sum is zero based).
Requiring now that the first Chern class, $N$ of Eq.~\eqref{eq:N}, is
positive semi-definite, eliminates the equations with the
coefficients, $-C_0$, $C_{2}$, $\ldots$, $C_{2L}$, in the subset
$\{-1,0\}$, of which there are 
\beq
\sum_{n=0}^{L+1}\begin{pmatrix} L+1\\ n\end{pmatrix} = 2^{L+1}.
\eeq
We will however allow for all the coefficients to vanish, of which
there is exactly one equation and it is the Laplace vortex equation.
Summing up, the final result is thus
\beq
HF(L) = 3^{L+1} - 2^{L+1} + 1,
\label{eq:HF}
\eeq
which is the Hilbert function in the counting of equations.
We can now write down the Hilbert series, which reads
\begin{align}
H(t) = \sum_{L=-1}^\infty HF(L) t^L
&= \sum_{L=-1}^\infty (3^{L+1} - 2^{L+1} + 1)t^L\non
&= \frac{1}{t}\left(\frac{1}{1-3t} - \frac{1}{1 - 2t} + \frac{1}{1-t}\right).
\end{align}
Expanding the first few terms of the Hilbert series yields
\beq
H(t) = \frac{1}{t} + 2 + 6t + 20t^2 + 66t^3 + 212t^4 + 666t^5 + \mathcal{O}(t^6),
\eeq
from which we can read off how many vortex equations there are at each
level, $L$.
Having no right hand side $L=-1$ (in this notation), gives only one
equation $HF(-1)=1$, namely the Laplace-vortex equation; allowing for
one coefficient ($L=0$), there is additionally the possibility of the
Bradlow vortex equation \cite{Manton:2016waw} and hence $HF(0)=2$;
allowing for two coefficients ($L=1$), there are now four more vortex
equations: the Taubes or hyperbolic, the Popov, the Jackiw-Pi and the
Ambj{\o}rn-Olesen vortex equations \cite{Manton:2016waw} and hence
$HF(1)=6$.
The $L=2$ case is the next step, which contains the relativistic
Chern-Simons vortex equation and is the class of vortex equations that
we will study in this paper.
In this case there are $HF(2)=20$ vortex equations as promised. 

Only for the case $L=1$, yielding six vortex equations, it is possible
to normalize all coefficients to $\pm 1$ if they are nonvanishing.
In the general case, there are $L$ real coefficients but only two ways
of absorbing constants (see below for details).
Therefore, we will in general have at most
\beq
W(L) \leq L-1,
\eeq
families of equations, meaning that there are $W(L)$ real parameters
in the equation whose nonvanishing magnitude can be varied (but not
set to zero).

\section{Classification of the vortex equations}\label{sec:types}

In this paper, we will study the vortex equation in the class of
equations that can be put in the following form
\beq
\Delta_{g_0}u 
+\frac{\delta_D}{\Omega_0}
= -C_0 + C_2 e^{2u} + C_4 e^{4u}.
\label{eq:vtxeq}
\eeq
The left-hand side is the magnetic flux and the right hand side is a
polynomial in the Higgs field squared ($e^{2u}$).
Since the magnetic flux is a real quantity and so is the vortex field
$u:\mathbb{C}\to\mathbb{R}$, all three coefficients must be real
valued $C_{0,2,4}\in\mathbb{R}$.
This class of vortex equations is the next order, compared to the five
equations considered in Ref.~\cite{Manton:2016waw}. 

The first classification that we will make is to count the number of
nonvanishing vortex coefficients (VC) and we will denote the type as
type $m$, with $m$ being the number of nonvanishing VCs. 

\subsection{Type 0}

No nonvanishing VCs reduces the vortex equation to the simplest
possible equation:
\beq
-4\p_z\p_{\zb} u
+\delta_D
= 0,
\label{eq:vtxeq_Laplace}
\eeq
which may be called the Laplace vortex equation.
This equation is of course independent of $L$.
The solutions depend on the choice of base geometry and the
boundary conditions. Suitability of the boundary conditions depend on
whether $M_0$ is compact or not.

A simple class of solutions can be constructed out of holomorphic
functions with singularities at $\{z_i\}$:
\beq
u = \sum_{i=1}^Nc_i\log|z - z_i|,
\eeq
with $c_i>0$ positive coefficients.

\subsection{Type I}

One nonvanishing VC means that there are $L+1$ vortex equations in
this class and for $L=2$ this implies that there are 3 equations:
$C_0<0$, $C_2>0$ and $C_4>0$, which correspond to the Bradlow equation
\cite{Manton:2016waw}
\beq
\Delta_{g_0} u
+\frac{\delta_D}{\Omega_0}
= 1,
\label{eq:vtxeq_Bradlow}
\eeq
the Jackiw-Pi equation \cite{Jackiw:1990tz}
\beq
\Delta_{g_0} u
+\frac{\delta_D}{\Omega_0}
= e^{2u},
\label{eq:vtxeq_Jackiw-Pi}
\eeq
and a new equation which we may call \type{I}{4}{+}:
\beq
\Delta_{g_0} u
+\frac{\delta_D}{\Omega_0}
= e^{4u}.
\label{eq:vtxeq_I4}
\eeq
The reason for the signs of the coefficients is that the magnetic flux
is positive ($N>0$) and $u:\mathbb{C}\to\mathbb{R}$ is a real field
and hence $e^{2u}$ is non-negative.
For the latter two equations, the coefficients $C_2$ and $C_4$ can be
set to unity, respectively, by a rescaling in
$u\to u-\tfrac12\log|C_2|$ and $u\to u-\tfrac14\log|C_4|$.
For the Bradlow equation, the coefficient can be rescaled
to unity by a coordinate transformation $z\to|C_0|^{-\frac12}z$
(the 2-dimensional delta function picks up the same factor as
$\p_z\p_{\zb}$ under the local coordinate rescaling).
Notice that the coordinate rescaling changes the conformal factor as
follows
\beq
\Omega_0 = \frac{4}{(1+K_0|z|^2)^2}
\to\frac{4}{\left(1+K_0|C_0|^{-1}|z|^2\right)^2}
=\frac{4}{(1+K_0'|z|^2)^2},
\eeq
with $K_0'$ the constant curvature of $M_0$ in the new coordinates. 
In order not to clutter the notation, we will simply drop the primes
of $K_0$ upon rescaling the coordinates by $|C_0|^{-\frac12}$.

\subsection{Type II}

The type of vortex equation with two nonvanishing VCs can for $L>1$ be
further classified into which of the coefficients are nonvanishing:
\type{II}{02}{}, \type{II}{04}{} and \type{II}{24}{}, which we will
discuss in turn below.
Using a scaling argument, the two nonvanishing coefficients can take
the values $\{-1,1\}$ which gives a total of $2^2$ vortex equations,
and the positive flux condition eliminates the choice of both of them
being $-1$.\footnote{To put the positiveness of the VCs on the same
  footing, we will talk about the signs of $-C_0$, $C_2$, and $C_4$.}
Since there are $\binom{L+1}{2}$ ways to set $L-1$ of the $L+1$
coefficients to zero, the total number of type II vortex equations is
thus: 
\beq
\binom{L+1}{2}(2^2-1)=\frac32L(L+1).
\eeq

\subsubsection{\texorpdfstring{\Type{II}{02}{}}{Type II02}}

This type of vortex equation was already covered in
Ref.~\cite{Manton:2016waw}
\beq
\Delta_{g_0} u
+\frac{\delta_D}{\Omega_0}
= -C_0 + C_2 e^{2u}.
\label{eq:vtxeq_II02}
\eeq
The scaling argument combining the shift in $u$ and the scaling of the
local coordinate $z$, yields $u\to u-\frac12\log|C_2/C_0|$ and
$z\to|C_0|^{-\frac12}z$, after which the Gauss curvature becomes
$K_0'=K_0|C_0|^{-1}$ (but we will drop the prime as usual) and the
only possible values for the coefficients $C_0$ and $C_2$ are:
$(C_0,C_2)=(-1,-1)$ for Taubes vortices, $(C_0,C_2)=(1,1)$ for Popov
vortices and $(C_0,C_2)=(-1,1)$ for Ambj{\o}rn-Olesen vortices.

\subsubsection{\texorpdfstring{\Type{II}{04}{}}{Type II04}}

The next type of vortex equation is new, but is conceptually similar
to the ones in the previous subsubsection
\beq
\Delta_{g_0} u
+\frac{\delta_D}{\Omega_0}
= -C_0 + C_4 e^{4u}.
\label{eq:vtxeq_II04}
\eeq
We will see in the next section that it is integrable, in the same way
as the Taubes, Popov and Ambj{\o}rn-Olesen vortices are, but with
different geometric interpretation.
The scaling argument is similar to the one above and is given by
$u\to u-\frac14\log|C_4/C_0|$ and $z\to |C_0|^{-\frac12}z$.
The possible values of the coefficients are thus $(C_0,C_4)=(-1,-1)$
for \type{II}{04}{--} vortices, $(C_0,C_4)=(1,1)$ for
\type{II}{04}{++} vortices and $(C_0,C_4)=(-1,1)$ for
\type{II}{04}{-+} vortices.

These vortex equations can be thought of as generalizations of the
Taubes, the Popov and the Ambj{\o}rn-Olesen vortex equations with the
Higgs field squared replaced by a quartic power of the Higgs field.

\subsubsection{\texorpdfstring{\Type{II}{24}{}}{Type II24}}

This type of vortex equation is known from the relativistic BPS
Chern-Simons theory \cite{Jackiw:1990aw} and takes the form
\beq
\Delta_{g_0} u
+\frac{\delta_D}{\Omega_0}
= C_2 e^{2u} + C_4 e^{4u}.
\label{eq:vtxeq_II24}
\eeq
This time $C_0=0$ so the rescaling is done as
$u\to u-\frac12\log|C_2/C_4|$ and $z\to|C_4|^{\frac12}|C_2|^{-1}z$. 
The possible coefficients, keeping in mind that the right-hand side of
the vortex equation must be able to have a positive integral, are:
$(C_2,C_4)=(1,1)$ for the \type{II}{24}{++} vortices,
$(C_2,C_4)=(1,-1)$ for the relativistic BPS Chern-Simons vortices and
$(C_2,C_4)=(-1,1)$ for the \type{II}{24}{-+} vortices. 

The \type{II}{24}{++} and \type{II}{24}{-+} vortex equations can be
thought of as being to the Ambj{\o}rn-Olesen and Popov vortex
equations what the Chern-Simons vortex equations are to the Taubes
equation.

\subsection{Type III}

Finally, we have the most complicated type of vortex equations for
which none of the 3 VCs vanish
\beq
\Delta_{g_0} u
+\frac{\delta_D}{\Omega_0}
= -C_0 + C_2 e^{2u} + C_4 e^{4u}.
\label{eq:vtxeq_III}
\eeq
It is a choice which VCs to scale to unity; we choose $C_0$ and
$C_2$ and hence we perform a rescaling by shifting
$u\to u-\frac12\log|C_2/C_0|$ and $z\to|C_0|^{-\frac12}z$ obtaining
\beq
\Delta_{g_0} u
+\frac{\delta_D}{\Omega_0}
= -\sign(C_0) + \sign(C_2) e^{2u} + C_4' e^{4u},
\label{eq:vtxeq_III_rescaled}
\eeq
where $C_4'=\frac{C_4|C_0|}{|C_2|^{2}}\neq 0$ and after the rescaling
we still have $\sign(C_4')=\sign(C_4)$.
We can now drop the prime on $C_4$ and recall that $C_0$ and $C_2$ can
be only $\pm1$.
There are now two possibilities: either $C_4>0$ or $C_4<0$, which we
will consider in turn below.

The number of type III vortex equations can be calculated as follows.
There are now 3 nonvanishing coefficients, which we classify according
to their sign, giving $2^3$ vortex equations.
The positive flux condition eliminates again one of these, since all
coefficients cannot be negative. For general $L$, there are now
$\binom{L+1}{3}$ ways of setting $L-2$ of $L+1$ coefficients to zero,
yielding a total of
\beq
\binom{L+1}{3}(2^3-1) = \frac76(L-1)L(L+1),
\eeq
vortex equations of type III. 

\subsubsection{\texorpdfstring{\Type{III}{024}{±±+}}{Type III024±±+}}

There are 4 vortex equations that obey the positive flux condition for
$C_4>0$:
the \type{III}{024}{--+} vortex equation with $(C_0,C_2)=(-1,-1)$,
the \type{III}{024}{-++} vortex equation with $(C_0,C_2)=(-1,1)$,
the \type{III}{024}{+-+} vortex equation with $(C_0,C_2)=(1,-1)$ and
the \type{III}{024}{+++} vortex equation with $(C_0,C_2)=(1,1)$.

\subsubsection{\texorpdfstring{\Type{III}{024}{±±-}}{Type III024±±-}}

There are only 3 vortex equations that obey the positive flux
condition for $C_4<0$:
the \type{III}{024}{---} vortex equation with $(C_0,C_2)=(-1,-1)$,
the \type{III}{024}{-+-} vortex equation with $(C_0,C_2)=(-1,1)$ and
the \type{III}{024}{++-} vortex equation with $(C_0,C_2)=(1,1)$.

\subsection{Summary of types}

We will now summarize the types of vortex equations in the following
table.
\begin{table}[!htp]
\begin{center}
\begin{tabular}{llrrr}
  type & name & $C_0$\! & $C_2$\! & $C_4$\!\\
  \hline\hline
  type 0 & ``Laplace'' & $0$ & $0$ & $0$\\
  \hline
  \type{I}{0}{-} & Bradlow & $-1$ & $0$ & $0$\\
  \type{I}{2}{+} & Jackiw-Pi & $0$ & $+1$ & $0$\\
  \type{I}{4}{+} & -- & $0$ & $0$ & $+1$\\
  \hline
  \type{II}{02}{--} & Taubes & $-1$ & $-1$ & $0$\\
  \type{II}{02}{++} & Popov & $+1$ & $+1$ & $0$\\
  \type{II}{02}{-+} & Ambj{\o}rn-Olesen & $-1$ & $+1$ & $0$\\
  \hline
  \type{II}{04}{--} & -- & $-1$ & $0$ & $-1$\\
  \type{II}{04}{++} & -- & $+1$ & $0$ & $+1$\\
  \type{II}{04}{-+} & -- & $-1$ & $0$ & $+1$\\
  \hline
  \type{II}{24}{+-} & Chern-Simons & $0$ & $+1$ & $-1$\\
  \type{II}{24}{-+} & -- & $0$ & $-1$ & $+1$\\
  \type{II}{24}{++} & -- & $0$ & $+1$ & $+1$\\
  \hline
  \type{III}{024}{--+} & -- & $-1$ & $-1$ & $+1$\\
  \type{III}{024}{-++} & -- & $-1$ & $+1$ & $+1$\\
  \type{III}{024}{+-+} & -- & $+1$ & $-1$ & $+1$\\
  \type{III}{024}{+++} & -- & $+1$ & $+1$ & $+1$\\
  \hline
  \type{III}{024}{---} & -- & $-1$ & $-1$ & $-1$\\
  \type{III}{024}{-+-} & -- & $-1$ & $+1$ & $-1$\\
  \type{III}{024}{++-} & -- & $+1$ & $+1$ & $-1$\\
\end{tabular}
\caption{The classification of vortex equations for $L=2$, i.e.~the
  class of twenty vortex equations described by Eq.~\eqref{eq:vtxeq}.
}
\label{tab:classification}
\end{center}
\end{table}

For $L=2$, we only have four major types of vortex equations in the
classification, namely type 0, type I, type II and type III; summing
up the number of different vortex equation in each type, we get
\beq
\left.1 + (L+1) + \frac32L(L+1) + \frac76(L-1)L(L+1)\right|_{L=2}
=20,
\eeq
as expected.
As a consistency check, we can contemplate the situation with $L+1$
being general and not equal to two, for which there would now be $L+2$
major types (the first being type 0).
The counting can easily be seen to be done by
\beq
1+\sum_{i=1}^{L+1}\binom{L+1}{i}(2^i-1)
= 3^{L+1} - 2^{L-1} + 1,
\eeq
as it should, see Eq.~\eqref{eq:HF}.

\section{Integrability}\label{sec:integrability}

We will now turn to the question of integrability of a subset of the
vortex equations, reviewing the known ones and presenting the new
integrable vortex equations.

\subsection{\texorpdfstring{\Type{II}{02}{}}{Type II02}}\label{sec:int_typeII02}

We will review the integrability in this type of vortex equations
following Ref.~\cite{Manton:2016waw}. Using the scaling arguments
described in the previous section, the coefficients $C_0=\pm1$ and
$C_2=\pm'1$ and the vortex equation is given in
Eq.~\eqref{eq:vtxeq_II02}.
The strategy of the integrability utilized in this type of equation is
to remove the constant term $C_0$ by finding a suitable geometry for
$M_0$ and then integrate the remaining vortex field \`a la Liouville.
In particular, we note that
\beq
\Delta_{g_0}u = -\frac{4}{\Omega_0}\p_z\p_{\zb}u
= -\frac{4}{\Omega_0}\p_z\p_{\zb}u'
+ \frac{2}{\Omega_0}\p_z\p_{\zb}\log\Omega_0
= \Delta_{g_0}u' - K_0,
\label{eq:K0shift}
\eeq
with
\beq
u = u' - \frac12\log\Omega_0.
\label{eq:ushift}
\eeq
It is thus clear that a constant curvature manifold $M_0$ with Gauss
curvature $K_0=C_0$ effectively eliminates the constant term as well
as the conformal factor from the vortex equation:
\beq
-4\p_z\p_{\zb} u'
+\delta_D
= C_2 e^{2u'}.
\label{eq:vtxeq_II02_on_geom}
\eeq
Exponentiating Eq.~\eqref{eq:ushift} yields
\beq
e^{2u'} = \Omega_0 e^{2u} = \Omega_2,
\eeq
so indeed the vortex equation in the form of
Eq.~\eqref{eq:vtxeq_II02_on_geom} can be interpreted as a constant
Baptista curvature \eqref{eq:K2}:
\begin{equation}
-\frac{2}{\Omega_2}\p_z\p_{\zb}\log\Omega_2
+\frac{\delta_D}{\Omega_2}
= K_2 +\frac{\delta_D}{\Omega_2}
=C_2,
\label{eq:constant_K2}
\end{equation}
with the exception of certain singularities at $\{z_i\}$ where the
Baptista metric is degenerate.

An elegant way of writing the vortex equation of this type, is to
notice the form of the Baptista curvature (away from vortex
singularities $\{z_i\}$) of Eq.~\eqref{eq:K2} can be rewritten as
\beq
\frac{\Omega_2}{\Omega_0}K_2 - K_0
= -K_0 + K_2 e^{2u} 
= - \frac{4}{\Omega_0}\p_z\p_{\zb}u,
\eeq
which is exactly the vortex equation \eqref{eq:vtxeq_II02} on
$M_0\backslash\{z_i\}$.
Equating the above equation with the vortex equation, yields the
Baptista equation
\beq
-K_0 + K_2 e^{2u} = -C_0 + C_2 e^{2u},
\eeq
or on Baptista form
\beq
\Omega_0(K_0 - C_0) = \Omega_2(K_2 - C_2),
\label{eq:Baptista}
\eeq
which is only defined in $M_0\backslash\{z_i\}$.
Integrability corresponds to the ``trivial'' case where each side of
the equation vanishes separately.

The constant curvature equation for $K_0=C_0=\pm1$ is simply solved by
\beq
\Omega_0 = \frac{4}{(1 + C_0|z|^2)^2},
\eeq
and likewise, so is the constant Baptista curvature condition or vortex equation
\eqref{eq:vtxeq_II02_on_geom}:
\beq
\Omega_2 = \frac{4}{(1 + C_2|z|^2)^2},
\eeq
but it bears no vortices.
A simple remedy is to use the chain rule, giving
\beq
\Omega_2 = e^{2u'} = \frac{4}{(1 + C_2|f(z)|^2)^2}\left|\frac{\d f}{\d z}\right|^2,
\eeq
where the ramification points of $f(z)$ (i.e.~the points where
$f'(z)=0$) correspond to vortex positions or vortex centers.
The original Higgs field is related to the above solution via the
solution to the constant curvature solution on $M_0$ as
\beq
|\phi|^2 = e^{2u} = \frac{\Omega_2}{\Omega_0}
= \frac{(1 + C_0|z|^2)^2}{(1 + C_2|f(z)|^2)^2}\left|\frac{\d f}{\d z}\right|^2.
\label{eq:phisq_int_typeII02}
\eeq
This expression lends an easy way to calculate $\phi$ in a certain
gauge (holomorphic gauge), but the gauge can be chosen arbitrarily, of
course.

\subsection{\texorpdfstring{\Type{II}{04}{}}{Type II04}}

The vortex equation is now given by Eq.~\eqref{eq:vtxeq_II04} with
$C_0=\pm1$ and $C_4=\pm'1$ which are found by using the scaling
arguments of Sec.~\ref{sec:types}; the integrability in this type of
vortex equation is to the best of our knowledge new.
They can be thought of as the higher-order generalizations of the
Taubes, Ambj{\o}rn-Olesen and Popov vortex equations with $e^{2u}$
replaced by $e^{4u}$.
The strategy for uncovering integrability is the same as in the
\type{II}{02}{} vortices, but we have to modify the change of
variables as
\beq
\Delta_{g_0}u = \Delta_{g_0}u' - \frac{K_0}{2},
\label{eq:Deltau_change4}
\eeq
with
\beq
u = u' - \frac14\log\Omega_0.
\eeq
The change of the fraction in the above change of variables from $u$
to $u'$ is dictated by the exponentiation
\beq
e^{4u'} = \Omega_0 e^{4u} = \Omega_4,
\eeq
where we want a linear dependence on $\Omega_0$ for not mocking up
integrability.
Interestingly, this changes the coefficient in
Eq.~\eqref{eq:Deltau_change4} from unity to a half.
Setting now $K_0=2C_0$ eliminates the constant term in the vortex
equation and leaves us with
\beq
-4\p_z\p_{\zb}u'
+\delta_D
= C_4 e^{4u'}.
\eeq
This equations can now be interpreted as a constant generalized
Baptista curvature \eqref{eq:K2n}:
\begin{equation}
-\frac{1}{\Omega_4}\p_z\p_{\zb}\log\Omega_4
+\frac{\delta_D}{\Omega_4}
= \frac{K_4}{2}
+\frac{\delta_D}{\Omega_4}
= C_4,
\end{equation}
with the exception of the singularities at $\{z_i\}$ where the
generalized Baptista metric is degenerate.
As in section \ref{sec:int_typeII02}, we recognize that the
generalized Baptista curvature \eqref{eq:K2n} (away from vortex
singularities and for $n=2$) can be written as
\beq
\frac{\Omega_4}{2\Omega_0} - \frac{K_0}{2}
= -\frac{K_0}{2} + \frac{K_4}{2}e^{4u}
= -\frac{4}{\Omega_0}\p_z\p_{\zb}u,
\eeq
which has the same form as the vortex equation \eqref{eq:vtxeq_II04}
on $M_0\backslash\{z_i\}$.
Equating the above equation with the vortex equation yields the
higher-order Baptista equation
\beq
-\frac{K_0}{2} + \frac{K_4}{2}e^{4u}
=-C_0 + C_4e^{4u},
\eeq
which can be written on Baptista form as
\beq
\Omega_0(K_0 - 2C_0) = \Omega_4(K_4 - 2C_4),
\label{eq:Baptista4}
\eeq
which is again only defined on $M_0\backslash\{z_i\}$.
Integrability corresponds to the ``trivial'' solution of setting each
side of the above equation to zero.

The constant curvature solution for $K_0=2C_0=\pm2$ thus reads
\beq
\Omega_0 = \frac{4}{(1+2C_0|z|^2)^2},
\eeq
whereas the constant generalized Baptista curvature is solved by
\beq
\Omega_4 = e^{4u'}
= \frac{4}{(1+2C_4|f(z)|^2)^2}\left|\frac{\d f}{\d z}\right|^2,
\eeq
where ramification points of $f(z)$ correspond to vortex positions.
The original Higgs field is now related to the solution by
\beq
|\phi|^2 = e^{2u} = \sqrt{\left|\frac{\Omega_4}{\Omega_0}\right|}
=\frac{|1+2C_0|z|^2|}{|1+2C_4|f(z)|^2|}\left|\frac{\d f}{\d z}\right|.
\label{eq:phisq_int_typeII04}
\eeq
This solution does not lend an easy choice of gauge that fixes the
phase in terms of $f(z)$.

\subsection{Type I}

\subsubsection{Bradlow}

Integrability of the Bradlow or \type{I}{0}{-} equation was discussed
in Ref.~\cite{Manton:2016waw} on hyperbolic space and further in
Ref.~\cite{Gudnason:2017jsn} on nontrivial geometries with
nonconstant curvature.
As noticed in Ref.~\cite{Manton:2016waw}, it is possible to simply set
$C_2:=0$ in the solution \eqref{eq:phisq_int_typeII02} which yields a
class of solutions on the hyperbolic plane.

\subsubsection{Jackiw-Pi}

Integrability of the Jackiw-Pi or \type{I}{2}{+} equation was discussed
in Ref.~\cite{Manton:2016waw}. On flat space $(K_0=0)$ it is possible
to simply set $C_0:=0$ in the solution \eqref{eq:phisq_int_typeII02}
which yields a class of solutions on $\mathbb{R}^2$ or $T^2$.

\subsubsection{\texorpdfstring{\Type{I}{4}{}}{Type I4}}

Integrability of the \type{I}{4}{+} equation is, to the best of our
knowledge, new. The solution can be found by simply setting $C_0:=0$
in Eq.~\eqref{eq:phisq_int_typeII04} which yields a class of solutions
on $\mathbb{R}^2$ or $T^2$.
This equation can be thought of as the higher-order generalization of
the Jackiw-Pi vortex equation with $e^{2u}$ replaced by $e^{4u}$. 

\subsection{Summary of integrable vortex equations}

We now summarize the integrable vortex equations in
Tab.~\ref{tab:integrable}.
\begin{table}[!htp]
\begin{center}
\begin{tabular}{llrrrccc}
  type & name & $C_0$\! & $C_2$\! & $C_4$\! & $M_0$ & $M_2$ & $M_4$\\
  \hline\hline
  type 0 & ``Laplace'' & $0$ & $0$ & $0$ & $\mathbb{R}^2$ & $\mathbb{R}^2$ & $\mathbb{R}^2$\\
  \hline
  \type{I}{0}{-} & Bradlow & $-1$ & $0$ & $0$ & $\mathbb{H}^2$ & $\mathbb{R}^2$ & $\mathbb{R}^2$\\
  \type{I}{2}{+} & Jackiw-Pi & $0$ & $+1$ & $0$ & $\mathbb{R}^2$ & $\mathbb{S}^2$ & --\\
  \type{I}{4}{+} & -- & $0$ & $0$ & $+1$ & $\mathbb{R}^2$ & -- & $\mathbb{S}^2$\\
  \hline
  \type{II}{02}{--} & Taubes & $-1$ & $-1$ & $0$ & $\mathbb{H}^2$ & $\mathbb{H}^2$ & --\\
  \type{II}{02}{++} & Popov & $+1$ & $+1$ & $0$ & $\mathbb{S}^2$ & $\mathbb{S}^2$ & --\\
  \type{II}{02}{-+} & Ambj{\o}rn-Olesen & $-1$ & $+1$ & $0$ & $\mathbb{H}^2$ & $\mathbb{S}^2$ & --\\
  \hline
  \type{II}{04}{--} & -- & $-1$ & $0$ & $-1$ & $\mathbb{H}^2$ & -- & $\mathbb{H}^2$\\
  \type{II}{04}{++} & -- & $+1$ & $0$ & $+1$ & $\mathbb{S}^2$ & -- & $\mathbb{S}^2$\\
  \type{II}{04}{-+} & -- & $-1$ & $0$ & $+1$ & $\mathbb{H}^2$ & -- & $\mathbb{S}^2$\\
\end{tabular}
\caption{The integrable subset of vortex equations for $L=2$, i.e.~among the 
  class of twenty vortex equations described by Eq.~\eqref{eq:vtxeq}.
  The last three columns display the geometries of the (constant
  curvature) manifolds $M_0$, Baptista manifolds $M_2$ and
  higher-order Baptista manifolds $M_4$.
  The vanishing constant curvature manifolds are denoted by
  $\mathbb{R}^2$, but they can equally well be tori: $T^2$ etc.
  The -- denotes manifolds that do not have a constant curvature and
  hence a more complicated geometry.
  The Baptista manifolds have conical singularities at the vortex
  zeros $\{z_i\}$.
}
\label{tab:integrable}
\end{center}
\end{table}

\section{Volumes}\label{sec:volumes}

For Taubes vortices, the integral of the vortex equation leading to a
bound between the number of vortices and surface area of $M_0$ was
considered by Bradlow in Ref.~\cite{Bradlow:1990ir}.
Here we will make similar considerations for the class of vortex
equations at hand.
In particular, by integrating Eq.~\eqref{eq:BPS2} on $M_0$, we get
\beq
\int_{M_0} F = 2\pi N = \int_{M_0}P(|\phi|^2)\omega_0
= \int_{M_0} P(|\phi|^2)\Omega_0\;\d x^1\wedge\d x^2,
\eeq
where we have used the first Chern class of the line bundle
\eqref{eq:N} being the winding number $N$ or simply the vortex
number.
Inserting the vortex polynomial \eqref{eq:vtxeq}, we have
\begin{align}
2\pi N &= -C_0\int_{M_0} \omega_0
+ C_2\int_{M_0} e^{2u} \omega_0
+ C_4\int_{M_0} e^{4u} \omega_0\non
&= -C_0 \Vol(M_0) + C_2 \Vol(M_2) + C_4 \Vol(M_4),
\label{eq:int_vtxeq}
\end{align}
that is the winding number $N$ is related to the volumes of the
manifolds $M_0$ (the Riemann surface), $M_2$ and $M_4$, where the
latter two manifolds are $M_0$ equipped with the conformally
transformed metrics $g_2$ and $g_4$, respectively.

Let us consider the case where $M_0$ is smooth and compact with
genus $\mathfrak{g}_0$. Then by the Gauss-Bonnet theorem, we have
\beq
\int_{M_0} K_0\omega_0
= \int_{M_0} K_0\Omega_0\;\d x^1\wedge\d x^2
= 2\pi\chi_E(M_0)
= 4\pi(1 - \mathfrak{g}_0),
\eeq
where $\chi_E(M_0)$ is the Euler characteristic of $M_0$. 
Similarly, for the Baptista metric we have
\beq
\int_{M_0} K_2e^{2u}\omega_0
= \int_{M_0} K_2\Omega_2\;\d x^1\wedge\d x^2
= 4\pi(1 - \mathfrak{g_0}).
\eeq
Now relating it to the Baptista area, $\Vol(M_2)$, using
Eq.~\eqref{eq:constant_K2} we have
\begin{equation}
C_2\int_{M_0} e^{2u}\omega_0
= C_2\Vol(M_2)
= \int_{M_0} K_2e^{2u}\omega_0 + 2\pi N
= 2\pi(2 - 2\mathfrak{g}_0 + N).
\end{equation}
The additional term $2\pi N$ comes from the $N$ points where the
Baptista metric is degenerate.

\subsection{Integrable vortices of \texorpdfstring{\type{II}{02}{}, \type{I}{0}{} and \type{I}{2}{}}{typeII02, typeI0 and typeI2}}

With the above integrals in hand, we can now integrate the Baptista
equation \eqref{eq:Baptista}, valid for the integrable vortex
equations of \type{II}{02}{}, \type{I}{0}{} and \type{I}{2}{},
obtaining
\beq
4\pi(1-\mathfrak{g}_0) - C_0\Vol(M_0)
= 2\pi(2 - 2\mathfrak{g}_0 + N) - C_2\Vol(M_2).
\eeq
The Euler characteristic times $2\pi$ cancels out, leaving us with
\beq
C_2\Vol(M_2) = C_0\Vol(M_0) + 2\pi N,
\label{eq:Baptista_vol_relation}
\eeq
and the values of $C_0$ and $C_2$ for the mentioned types of vortex
equations are given in Tab.~\ref{tab:classification}.
This latter equation is in fact true also on noncompact $M_0$, which
can be seen from the integral of the vortex equation
\eqref{eq:int_vtxeq} with $C_4:=0$.
In that case the volumes would diverge and should be considered as
areas of a disc $B\subset M_0$ in the limit of the radius being sent
to infinity.

Since the areas are non-negative, the signs of $C_0$ and $C_2$
determine whether there is an upper or lower bound on the number of
vortices $N$.
In particular, for the Taubes equation ($C_0=C_2=-1$) we have
\beq
\Vol(M_2) = \Vol(M_0) - 2\pi N,
\eeq
indicating that the Baptista volume is smaller than the volume (area)
of $M_0$ \cite{Manton:2016waw}.
Since $\Vol(M_2)\geq 0$ Bradlow's bound follows
\beq
N \leq \frac{\Vol(M_0)}{2\pi}.
\eeq
For Popov vortices $C_0=C_2=1$ means that the Baptista volume is
bigger than the volume (area) of $M_0$:
\beq
\Vol(M_2) = \Vol(M_0) + 2\pi N.
\eeq
For the Ambj{\o}rn-Olesen vortices, $C_0=-1$ and $C_2=1$, so
\beq
\Vol(M_2) = -\Vol(M_0) + 2\pi N,
\eeq
yielding a \emph{lower} bound on the vortex number
\beq
N \geq \frac{\Vol(M_0)}{2\pi}.
\eeq
For the Jackiw-Pi vortices, $C_2=1$ and $C_0=0$, and hence the
Baptista volume \emph{is} the vortex number (times $2\pi$):
\beq
\Vol(M_2) = 2\pi N.
\eeq
Similarly, for the Bradlow vortices, $C_0=-1$ and $C_2=0$, so the
vortex number is proportional to the volume (area) of $M_0$:
\beq
\Vol(M_0) = 2\pi N.
\eeq

\subsection{Integrable vortices of \texorpdfstring{\type{II}{04}{} and \type{I}{4}{}}{typeII04 and typeI2}}

We will now turn to the higher-order integrable vortex equations,
i.e.~those that involve the higher-order Baptista metric $g_4$.
Integrating now the higher-order Baptista equation
\eqref{eq:Baptista4}, valid for the integrable vortex equations of
\type{II}{04}{}, and \type{I}{4}{}, we obtain
\beq
4\pi(1-\mathfrak{g}_0) - 2C_0\Vol{M_0}
= 2\pi(2 - 2\mathfrak{g}_0 + 2N) - 2C_4\Vol(M_4),
\eeq
where the $4\pi N$ coming from the Gauss-Bonnet integral is due to the
conical excess of $4\pi$ per vortex zero (see the next section).
The Euler characteristic times $2\pi$ cancels out of the above
equation, yielding
\beq
C_4\Vol(M_4) = C_0\Vol(M_0) + 2\pi N,
\eeq
so although the curvatures $K_0=2C_0$ and $K_4=2C_4$ have a factor of
two in their relations to the VCs, this equation is formally identical
to Eq.~\eqref{eq:Baptista_vol_relation}
and the values of $C_0$ and $C_4$ are given in Tab.~\ref{tab:classification}.
This equation is again true on noncompact $M_0$, which can be seen
from the integral of the vortex equation \eqref{eq:int_vtxeq} with
$C_2:=0$.

We now have analogous relations for the higher-order vortex equations,
i.e.~since the volumes (areas) are non-negative, we have for the
\type{II}{04}{--} vortices
\beq
\Vol(M_4) = \Vol(M_0) - 2\pi N,
\eeq
yielding a Bradlow bound
\beq
N \leq \frac{\Vol(M_0)}{2\pi}.
\eeq
For the \type{II}{04}{++} vortices, the higher-order Baptista volume
is instead bigger than the volume (area) of $M_0$:
\beq
\Vol(M_4) = \Vol(M_0) + 2\pi N.
\eeq
For the \type{II}{04}{-+} vortices, we have
\beq
\Vol(M_4) = -\Vol(M_0) + 2\pi N,
\eeq
which like in the case of the Ambj{\o}rn-Olesen vortices yields a
lower bound on the vortex number
\beq
N \geq \frac{\Vol(M_0)}{2\pi}.
\eeq
Finally, for the \type{I}{4}{+} vortices, the higher-order Baptista
volume is related directly to the winding number
\beq
\Vol(M_4) = 2\pi N.
\eeq

\section{Vortex cones and deficits}\label{sec:deficits}

One may naively expect that the behavior of the field $u$ near a zero
$z_k\in\{z_i\}$ for some $k$ depends on the vortex polynomial
$P(e^{2u})$, but this is not so.
In fact, it suffices to know that the selfdual equation
\eqref{eq:BPS1} holds and that the vortex field satisfies any boundary
condition
\beq
\lim_{z\to z_0}e^{2u} > 0, \qquad
\forall z_0\in\p M_0,
\eeq
then the following proposition holds \cite{JT}.
Let $(A,\phi)$ be a smooth solution to the first order equation
\eqref{eq:BPS1}.
Then the set of zeros $\{z_i\}$, $i=1,\ldots,N$ is discrete and in some
neighborhood of each $z_k\in\{z_i\}$,
\beq
\phi(z,\zb) = (z-z_k)^{N_k}h_k(z,\zb),
\label{eq:phi_local_expansion}
\eeq
where $N_k\in\mathbb{Z}_+$ is the order of the zero and $h_k(z,\zb)$ is
$C^\infty$ and nonvanishing on the neighborhood.
The proof utilizes the $\bar{\p}$-Poincar\'e lemma (for the proof see
Ref.~\cite[page 5]{GH}), which states that 
$\i A(z,\zb)$ is a $C^\infty$ function on a closed disc
$B\subset\mathbb{R}^2$.
Then the differential equation
\beq
\p_{\zb}\beta(z,\zb) = \i A(z,\zb),
\eeq
has a $C^\infty$ solution $\beta$ in the interior of $B$:
\beq
\beta = \frac{1}{2\pi}\int_B\frac{A(\zeta,\bar\zeta)}{\zeta-z}\d\zeta\wedge\d\bar\zeta.
\eeq
Now using that $\eta(z)=e^{-\beta(z,\zb)}\phi(z,\zb)$ is a
holomorphic function
\beq
\p_{\zb}(e^{-\beta}\phi)
= e^{-\beta}\big(\p_{\zb}\phi - (\p_{\zb}\beta)\phi\big)
= e^{-\beta}(\p_{\zb} - \i A(z,\zb))\phi = 0,
\eeq
$\phi=e^{\beta}\eta$ is a product of a $C^\infty$ nonvanishing
function $e^\beta$ and a holomorphic function $\eta(z)$. Since the
zeros of a complex holomorphic function are discrete, a finite number
of zeros occur in any bounded set $B\subset\mathbb{R}^2$.
We can thus write
\beq
\eta(z) = (z-z_k)^{N_k}\eta_k(z),
\eeq
and in turn
\beq
h_k(z,\zb) = \eta_k(z)e^{\beta(z,\zb)}.
\eeq
Finally Eq.~\eqref{eq:phi_local_expansion} follows.

Now that we have established that Eq.~\eqref{eq:phi_local_expansion}
holds and that $h_k$ is nonvanishing near the zero $z_k\in\{z_i\}$ for
some $k$ and the multiplicity of the zero is $N_k$, we have
\cite{Manton:2016waw}
\beq
e^{2u} = |\phi|^2
= |h_k|^2|z-z_k|^{2N_k},
\eeq
and thus
\beq
\Omega_2 = \Omega_0e^{2u}
\simeq \Omega_0(|z_k|)|h_k(z_k,\zb_k)|^2|z-z_k|^{2N_k},
\eeq
in the vicinity of $z_k$.
The Baptista metric can then in the near vicinity of $z_k$ be written
in local coordinates $(z,\zb)$ as
\begin{align}
g_2 = \Omega_2\d z\d\zb 
&= \Omega_2\d(z-z_k)\d(\zb-\zb_k) \non
&\simeq \Omega_2(|z_k|)|h_k(z_k,\zb_k)|^2 r^{2N_k}(\d r^2 + r^2\d\theta^2) \non
&\simeq \Omega_2(|z_k|)|h_k(z_k,\zb_k)|^2 (\d\rho^2 + \rho^2\d\chi^2),
\end{align}
where in the second equality we have shifted the coordinate system to
place the origin at $z_k$, in the third equality we have changed to
polar coordinates $z-z_k=re^{\i\theta}$ and in the fourth equality we
have changed variables to
\beq
\rho=\frac{1}{N_k+1}r^{N_k+1}, \qquad
\chi = (N_k+1)\theta.
\eeq
Since $\theta\in[0,2\pi]$ the range of $\chi$ is given by
$\chi\in[0,2\pi(N_k+1)]$.
The geometry of the Baptista manifold is thus that of cones at the
vortex zeros $\{z_i\}$ with a deficit angle of $2\pi N_k$ at each zero
$z_k\in\{z_i\}$ with multiplicity $N_k$.
The cones are not flat cones, except in the case of Bradlow vortices,
as $C_2=0$ implies vanishing Baptista curvature $K_2=0$.
On the other hand, they are cones with constant curvature $K_2=C_2$,
except at the vortex zeros $\{z_i\}$.

For higher-order Baptista manifolds, in particular for the integrable
cases of \type{II}{04}{} and \type{I}{4}{}, the generalized Baptista
curvature $K_4=2C_4$ is also constant, but the local geometry is
different.

In particular, we have
\beq
\Omega_{2n} = \Omega_0e^{2nu}
\simeq \Omega_0(|z_k|)|h_k(z_k,\zb_k)|^{2n}|z-z_k|^{2nN_k},
\eeq
in the vicinity of $z_k$ and generalizing now to the $n$-th order
Baptista metric, which near the vicinity of $z_k$ can be written in
local coordinates $(z,\zb)$ as
\begin{align}
g_{2n} = \Omega_{2n}\d z\d\zb 
&= \Omega_{2n}\d(z-z_k)\d(\zb-\zb_k) \non
&\simeq \Omega_{2n}(|z_k|)|h_k(z_k,\zb_k)|^{2n} r^{2nN_k}(\d r^2 + r^2\d\theta^2) \non
&\simeq \Omega_{2n}(|z_k|)|h_k(z_k,\zb_k)|^{2n} (\d\rho^2 + \rho^2\d\chi^2),
\end{align}
where now we have defined
\beq
\rho = \frac{1}{n N_k + 1} r^{n N_k + 1},\qquad
\chi = (n N_k + 1)\theta.
\eeq
We can now readily see that the cone deficits of the higher-order
Baptista metrics becomes higher by a factor of $n$ and in particular
for $n=2$ (i.e.~the case of constant $K_4$ Baptista curvature), the
deficits are $4\pi N_k$ at each zero $z_k\in\{z_i\}$ with multiplicity
$N_k$.
For example for a single \type{II}{04}{} vortex, the geometry is
described by a constant higher-order Baptista curvature manifold with
cone deficits $4\pi$ at a single localized vortex zero.

\section{Vortex superposition and relations among vortex equations}\label{sec:superposition}

We will now consider some geometrically interesting consequences of
the Baptista equation, in particular, how vortex solutions can be
found iteratively, supposing that one knows how to find the solution
to a given vortex equation.
The main observation is that the Baptista equation \eqref{eq:Baptista}
is symmetric in the metrics $g_0$ and $g_2$; this fact can be
understood as the quantity on each side of the Baptista equation being
invariant under a degenerate conformal transformation of the metric
$\Omega_0\to e^{2u}\Omega_2$ \cite{Baptista:2012tx}.
Due to this transitivity property, the invariance can be used to find
new vortex solutions by ``adding'' vortices to known solutions.

\subsection{Vortices of \texorpdfstring{\type{II}{02}{}, \type{I}{0}{} and \type{I}{2}{}}{typeII02, typeI0 and typeI2}}

We will make the derivation in a slightly different way with respect
to the way it is presented in Ref.~\cite{Baptista:2012tx} (where only
the Taubes vortices were discussed).
Suppose we know a vortex solution $u_1$ which solves
\beq
\Delta_{g_0}u_1 + \frac{\delta_{D_1}}{\Omega_0}
=-C_0 + C_2 e^{2u_1}.
\label{eq:u1}
\eeq
Then consider the composite solution to the same equation on the same
background $(M_0,g_0)$:
\beq
\Delta_{g_0}(u_1 + u_2) + \frac{\delta_{D_1}+\delta_{D_2}}{\Omega_0}
=-C_0 + C_2 e^{2u_1 + 2u_2},
\eeq
which we can rewrite using the fact that $u_1$ solves
Eq.~\eqref{eq:u1} as
\beq
\Delta_{g_0}u_2 + \frac{\delta_{D_2}}{\Omega_0}
=-C_2 e^{2u_1} + C_2 e^{2u_1 + 2u_2},
\eeq
where it is understood that $D_1$ is the effective divisor containing
all the zeros of $u_1$ and $D_2$ is the same but for $u_2$.
Denoting by $\Omega_2^{(1)}:=e^{2u_1}\Omega_0$ and
$g_2^{(1)}:=\Omega_2^{(1)}\d z\d\zb$ the Baptista metric with the
specific solution $u_1$, we can write the above equation as
\beq
\Delta_{g_2^{(1)}}u_2 + \frac{\delta_{D_2}}{\Omega_2^{(1)}}
=-C_2 + C_2 e^{2u_2}.
\label{eq:u2}
\eeq
The observation made for the Taubes vortices in
Ref.~\cite{Baptista:2012tx}, is that the above equation is of the same
form as that of $u_1$, namely Eq.~\eqref{eq:u1}, but the geometric
interpretation of the vortex superposition is as follows:
The solution for $u_1$ is found on the background metric $g_0$, from
which the Baptista metric can be constructed, viz.~$g_2^{(1)}$, and
then the solution for $u_2$ can be found by solving exactly the same
equation, however with the background metric $g_0$ replaced by the
Baptista metric $g_2^{(1)}$ and $C_0$ replaced by $C_2$.
In the case of Taubes vortices studied in Ref.~\cite{Baptista:2012tx},
the signs of $C_0$ and $C_2$ are the same so the equation is exactly
the same\footnote{In Ref.~\cite{Baptista:2012tx}, a rescaling of the
  Baptista metric was performed so that the two vortex equations would
have the same constants in magnitude. We do not perform this rescaling
here, since we have already normalized the VCs and only the signs of
$C_0$ and $C_2$ remain. Furthermore, we do not allow for the conformal
transformation making the metric negative semi-definite.}.

A twist to the story is exactly that $C_0$ is replaced by $C_2$ in
Eq.~\eqref{eq:u2}: This has no consequence for the Taubes or the
Popov vortices, but for the Ambj{\o}rn-Olesen vortices, this implies
that the first solution is found for the Ambj{\o}rn-Olesen vortex 
equations, but an additional vortex added to the solution is found in
the background of the previous Baptista metric, but using now the
Popov vortex equation.
The same holds for the Jackiw-Pi vortices, a vortex can be added by
using the Baptista metric as the background metric and the Popov
vortex equation.
Bradlow vortices are added using the Laplace vortex equation. Since
the latter equation is homogeneous, it is not important whether the
background metric or the Baptista metric is used.

\subsection{Vortices of \texorpdfstring{\type{II}{04}{} and \type{I}{4}{}}{typeII04 and typeI4}}

We will now consider the analogous calculation as in the previous
subsection, but for the vortex equations involving a higher-order
Baptista metric.
Suppose we know a vortex solution $u_1$ which solves
\beq
\Delta_{g_0}u_1 + \frac{\delta_{D_1}}{\Omega_0}
= -C_0 + C_4 e^{4u_1}.
\label{eq:u1_4}
\eeq
Then consider the composite solution to the same equation on the same
background $(M_0,g_0)$:
\beq
\Delta_{g_0}(u_1 + u_2) + \frac{\delta_{D_1} + \delta_{D_2}}{\Omega_0}
= -C_0 + C_4^{4u_1 + 4u_2},
\eeq
which we can rewrite using Eq.~\eqref{eq:u1_4} as
\beq
\Delta_{g_0} u_2 + \frac{\delta_{D_2}}{\Omega_0}
= -C_4 e^{4u_1} + C_4 e^{2u_1 + 2u_2}.
\eeq
where it is understood that $D_1$ is the effective divisor containing
all the zeros of $u_1$ and $D_2$ is the same but for $u_2$.
Denoting by $\Omega_4^{(1)}:=e^{4u_1}\Omega_0$ and
$g_4^{(1)}:=\Omega_4^{(1)}\d z\d\zb$ the higher-order Baptista metric
with the specific solution $u_1$, we can write the above equation as
\beq
\Delta_{g_4^{(1)}}u_2 + \frac{\delta_{D_2}}{\Omega_4^{(1)}}
= -C_4 + C_4 e^{4u_2}.
\label{eq:u2_4}
\eeq
The solution for $u_1$ is found on the background metric $g_0$, from
which the higher-order Baptista metric can be constructed,
viz.~$g_4^{(1)}$, and then the solution for $u_2$ can be found by
solving the same type of vortex equation, however with the background
metric $g_0$ replaced by the Baptista metric $g_4^{(1)}$ and $C_0$
replaced by $C_4$.

For the \type{II}{04}{--} and \type{II}{04}{++} vortices $C_0$ and
$C_4$ have the same sign and are rescaled to unity, so the first
vortex $u_1$ and the addition $u_2$ are solved by the same vortex
equation.
However, for the \type{II}{04}{-+} and the \type{I}{4}{+} vortex
equation, the $u_1$ field is found by their corresponding equation
whereas the addition of the $u_2$ vortices is governed instead by the
\type{II}{04}{++} vortex equation and in the background of the
higher-order Baptista metric $g_4^{(1)}$.

\subsection{Vortices of \texorpdfstring{\type{III}{024}{}}{typeIII024}}\label{sec:rel_III024}

We will now consider the class of vortex equations of
\type{III}{024}{}
\beq
\Delta_{g_0} u
+\frac{\delta_D}{\Omega_0}
= -C_0 + C_2 e^{2u} + C_4 e^{4u},
\eeq
with $C_0\neq 0$, $C_2\neq 0$ and $C_4\neq 0$, which are not
integrable.
Let us first consider a superposition law for adding vortices to a
known solution, say $u_1$.
Then the composite solution $u_1+u_2$ on the same background
$(M_0,g_0)$ satisfies
\beq
\Delta_{g_0}(u_1 + u_2)
+\frac{\delta_{D_1}+\delta_{D_2}}{\Omega_0}
= -C_0 + C_2 e^{2u_1 + 2u_2} + C_4 e^{4u_1 + 4u_2},
\eeq
which using the equation for $u_1$ we can write as
\beq
-4\p_z\p_{\zb}u_2
+\delta_{D_2}
= C_2\Omega_2^{(1)}(e^{2u_2} - 1) + C_4\Omega_4^{(1)}(e^{4u_2} - 1),
\eeq
where $\Omega_{2n}^{(1)}:=e^{2n u_2}\Omega_0$, $n=1,2$ are Baptista
and higher-order Baptista metrics, respectively.
We can see that for both $C_2$ and $C_4$ nonvanishing, the
superposition law loses its nice geometric interpretation.

It is still possible to eliminate the constant factor, $-C_0$, from
the vortex equation by using a change of variables.
As in the integrable cases, we can write
\beq
u = u' - \frac{x}{2}\log\Omega_0, \qquad
x\in\mathbb{R},
\eeq
for which the vortex equation reduces to
\beq
\Delta_{g_0} u'
+\frac{\delta_D}{\Omega_0}
= \frac{C_2}{\Omega_0^x} e^{2u'} + \frac{C_4}{\Omega_0^{2x}} e^{4u'},
\eeq
if $xK_0=C_0$.
Two natural choices would be
\beq
-4\p_z\p_{\zb} u'
+\delta_D
&=& \sqrt{\Omega_0}C_2 e^{2u'} + C_4 e^{4u'},\label{eq:rel_III024_xhalf}\\
-4\p_z\p_{\zb} u'
+\delta_D
&=& C_2 e^{2u'} + \frac{C_4}{\Omega_0} e^{4u'},\label{eq:rel_III024_xunity}
\eeq
for $x=\frac12$ and $x=1$, respectively.
Hence, for the general case with $C_0\neq 0$, $C_2\neq 0$ and
$C_4\neq 0$, it is not possible to remove the constant term, $-C_0$,
by a change of variable without inducing a metric dependence into
either of the coefficients $C_2$ or $C_4$ (or both of them).
The vortex equation with $K_0=2C_0$ and $K_0=C_0$ are thus similar to
those with $C_0:=0$, but not equivalent due to the metric or space
dependence induced in the VCs, $C_2$ and $C_4$.

\subsection{Relations between \type{II}{24}{} and \type{II}{02}{} vortices}\label{sec:rel_II24_II02}

Last but finally, we will consider a geometric reinterpretation of the
vortex equations of \type{II}{24}{} as being vortex equations of
\type{II}{02}{} on the metric being the Baptista metric of the latter
equation.
More precisely, for the vortex equations of \type{II}{24}{}:
\beq
\Delta_{g_0} u
+\frac{\delta_D}{\Omega_0}
= C_2 e^{2u} + C_4 e^{4u},
\eeq
the observation is simply that using the Baptista metric
$\Omega_2=e^{2u}\Omega_0$, we can write the equation as
\beq
\Delta_{g_2} u
+\frac{\delta_D}{\Omega_2}
= C_2 + C_4 e^{2u},
\label{eq:rel_II24_II02_final}
\eeq
which is exactly the vortex equation of \type{II}{02}{} on the metric
containing the vortex solution itself, but with $C_2\to C_4$ and
$C_0\to C_2$.
The relation is geometrically interesting.

\subsection{Summary of relations between vortex equations}

\begin{figure}[!htp]
  \centering
  \includegraphics[width=\linewidth]{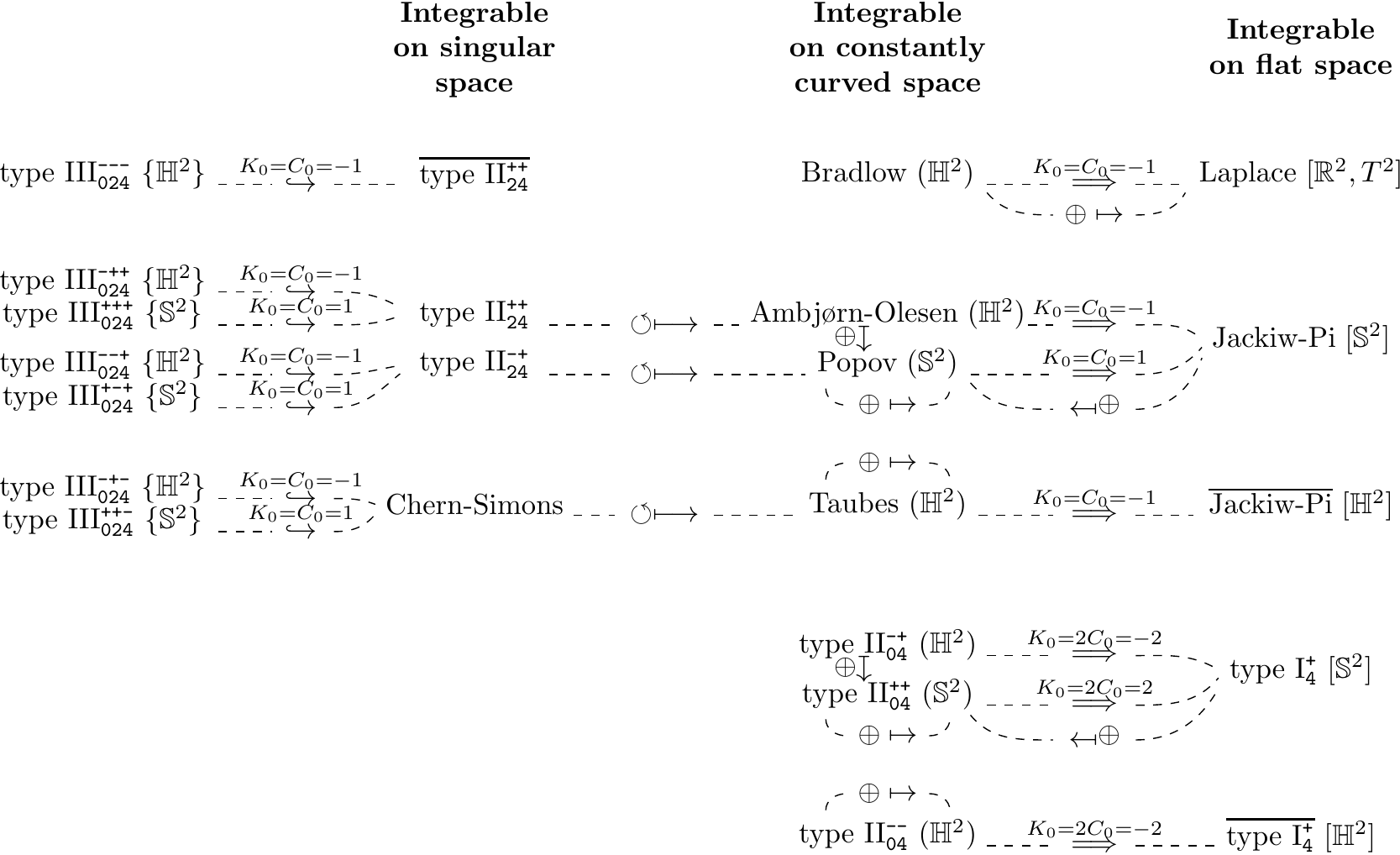}
  \caption{Map of relations between vortex equations for $L=2$
    (i.e.~20 vortex equations).
    From right, the columns are type I integrable, type II integrable
    and integrable on a singular space, with the exception of
    \type{II}{24}{++} which is not integrable and the Bradlow
    equation which is type I (and not type II) integrable.
    The different relations are equivalence by tuning the curvature of
    the background metric (denoted by $\Longrightarrow$),
    superposition law (denoted by $\oplus\mapsto$), the geometric
    interpretation of the vortex living on a manifold described by its
    own Baptista metric (denoted by $\circlearrowleft\longmapsto)$ and
    finally, a change of metric that relates vortex equations --
    but with metric dependent coefficients (denoted by $\hookrightarrow$).
    For more details, see the text.
  }
  \label{fig:relations}
\end{figure}

We will now summarize the relations amongst the vortex equations,
including the ones used for the integrable cases, the diagram of
relations is shown in Fig.~\ref{fig:relations}.
The figure shows a map of relations between vortex equations for $L=2$
(i.e.~20 vortex equations). The right-most column of equations are
type I integrable vortex equations and are hence integrable as they
are all the Liouville equation, with either sign, except for the
Laplace vortex equation, which is type 0 and is a harmonic problem.
The square brackets [] denote the geometry of the Baptista manifold
and the bar denotes the switch of the overall sign of the right-hand
side of the given equation.
The third column are type II integrable vortex
equations, except the Bradlow equation which is type I integrable.
The first kind of relation, denoted by the arrow $\Longrightarrow$ is
the equivalence between the vortex equations by choosing a specific
constant curvature for the background metric on the left-hand equation,
yielding the vortex equation on the right-hand side of the arrow on
flat space, see the third and fourth columns.
The round brackets () denote the geometry on which the vortex
equations are integrable.
The second kind of relation, denoted by the arrow $\oplus\mapsto$ are
superposition laws of adding vortices to a known solution, see the
third and fourth columns. 
The third kind of relation, denoted by the arrow
$\circlearrowleft\longmapsto$ is the map or geometric interpretation
that the vortex equation on the right-hand side has the background
metric replaced by the Baptista metric of the solution to the
equation, which is exactly the vortex equation on the left-hand side
of the arrow, see the second and third columns.
Finally, the last relation, denoted by the arrow $\hookrightarrow$ is
the relation that maps the vortex equation on the left-hand side to
the vortex equation on the right-hand side of the arrow, albeit with
at least one metric-dependent coefficient, by choosing a constant
background curvature, see the first and second columns.
The curly brackets \{\} denote the background geometry that reduces
the vortex equations via the latter relation.

\section{Further integrability and singular geometries}\label{sec:singular}

We may contemplate using the relations of Sec.~\ref{sec:rel_II24_II02}
between \type{II}{24}{} and \type{II}{02}{} vortices to find the
geometries that could bear an integrable vortex solution.
This is somewhat related to the Kazdan-Warner problem \cite{Kazdan:1974}.
The vortex equation \eqref{eq:rel_II24_II02_final} is integrable if we
tune the Baptista metric to have a constant curvature $-C_2$, that is:
\beq
\Omega_2 = \frac{4}{(1-C_2|z|^2)^2},
\eeq
for which the vortex equation reduces to
\beq
-4\p_z\p_{\zb}u' + \delta_D = C_4 e^{2u'},\qquad
u = u' - \frac12\log\Omega_2.
\eeq
The solution in terms of $u'$ is then simply
\beq
e^{2u'} = \frac{4}{(1+C_4|f(z)|^2)^2}\left|\frac{\d f}{\d z}\right|^2,
\eeq
and the original field $u$ is thus
\beq
e^{2u} = \frac{(1-C_2|z|^2)^2}{(1+C_4|f(z)|^2)^2}\left|\frac{\d f}{\d z}\right|^2.
\eeq
Since $\Omega_2=\Omega_0e^{2u}$, the background geometry $\Omega_0$
does not have a degenerate, but a singular metric
\beq
\Omega_0 = 4\frac{(1+C_4|f(z)|^2)^2}{(1-C_2|z|^2)^4}
\left|\frac{\d f}{\d z}\right|^{-2},
\eeq
with singularities at the ramification points, $f'(z)=0$, i.e.~the
vortex positions.
This class of singular geometries ``solve'' the Chern-Simons equation
for $C_2=1$, $C_4=-1$ as well as two further vortex equations, see
Tab.~\ref{tab:classification}. 
It might be possible to find some extensions of these metrics that are
better behaved.

\section{Discussion and outlook}\label{sec:discussion}

In this paper, we have considered vortex equations of the type that
naturally is derived from the Abelian Higgs model at critical
coupling, but generalized to having an arbitrary ``potential'' or
rather in our language an arbitrary vortex polynomial.
Using the positive flux condition, we have classified the possible
choices of vortex coefficients and found that for a vortex polynomial
that is at most quadratic in the Higgs field squared ($|\phi|^2)$,
corresponding to $L=2$, there are nineteen vortex equations other than
the (empty) Laplace vortex equation.
Although the classification of vortex equations becomes more
complicated with a drastically increasing number of vortex equations
as $L$ is increased, the number new of integrable vortex equations is
always four as $L$ is incremented by one: one new integrable equation
of \type{I}{2L}{+} and three new of \type{II}{0(2L)}{}.
We then establish the lower or upper bounds on the vortex number as a
function of the volume (area) of base manifold $(M_0,g_0)$, depending
on the type of vortex equation. There are also vortex equations for
which there are no such bound, for instance the Popov vortices.
Then we consider the geometry of the Baptista and higher-order
Baptista metrics and find that the Baptista manifolds always have
conical singularities at the vortex zeros, but with the conical
deficit increasing with the order of the Baptista metric.
We then consider the superposition rules and geometric interpretation
of finding new vortex solutions in the background of a known solution
for all the integrable types of vortices, generalizing the known
result of Baptista for Taubes vortices.
Finally, we contemplate further relations among the new vortex
equations, including an interpretation of \type{II}{24}{} vortices as
\type{II}{02}{} vortices in the Baptista background of their own
solutions. For example, a Chern-Simons vortex can be interpreted
geometrically as a Taubes vortex on the Baptista background of
itself. We have used this relation to find the singular geometry that
would be the solution to the Chern-Simons vortex.

This work leads to many questions for future work.
The first of which is what are the theories that these new vortex
equations could be derived from. 
For the $L=1$ case, i.e.~the five vortex equations of
Ref.~\cite{Manton:2016waw}, the theories behind the equations were
found in Ref.~\cite{Contatto:2017alh}.
A possible future direction of research would be to extend such
consideration to the new vortex equations (i.e.~$L=2$) studied in this
paper.
The theory behind the Chern-Simons equation is known
\cite{Jackiw:1990aw} and in fact it involves an extra field, that is,
the electric potential, in order to effectively get (at critical
coupling of that theory) the two coupled vortex equations included in
our class of equations.
Hence, the theories giving rise to the vortex equations contemplated
here may have to include further fields than simply a magnetic (gauge)
potential and a complex Higgs field for some of the equations.

It would be interesting to study the singular geometries that
are ``solutions'' to the Chern-Simons vortices and the other two
equations of \type{II}{24}{}, in particular whether the metric can be
regularized or made better behaved in some way.

It is clear that the four new integrable vortices of \type{I}{4}{+}, 
\type{II}{04}{--}, \type{II}{04}{++} and \type{II}{04}{-+} can be
readily generalized to even higher orders, i.e.~\type{I}{2n}{+}, 
\type{II}{0(2n)}{--}, \type{II}{0(2n)}{++} and \type{II}{0(2n)}{-+},
for which the vortex equation becomes
\beq
\Delta_{g_0}u +\frac{\delta_D}{\Omega_0}
= -C_0 + C_{2n}e^{2n u},
\eeq
the Baptista equation reads
\beq
\Omega_0(K_0 - n C_0) = \Omega_{2n}(K_{2n} - n C_{2n}),
\eeq
the Higgs field squared is
\beq
|\phi|^2 = \left|\frac{\Omega_{2n}}{\Omega_0}\right|^{\frac{1}{n}}
= \frac{|1 + n C_0|z|^2|^{\frac{2}{n}}}{|1 + n C_{2n}|f(z)|^2|^{\frac{2}{n}}}
\left|\frac{\d f}{\d z}\right|^{\frac{2}{n}}
\eeq
the conical deficits become $2\pi n N_k$ (for multiplicity $N_k$ of a
zero $z_k$) and the Bradlow bounds would be unchanged.
This establishes an infinite family of integrable vortex equations for
$n\in\mathbb{Z}_{>0}$ with coefficients $C_0=0$, $C_{2n}=1$ for
\type{I}{2n}{+}; $C_0=-1$, $C_{2n}=-1$ for \type{II}{0(2n)}{--};
$C_0=1$, $C_{2n}=1$ for \type{II}{0(2n)}{++} and $C_0=-1$, $C_{2n}=1$
for \type{II}{0(2n)}{-+}.
For $n=1$ these equations are four of the five vortex equations of
Ref.~\cite{Manton:2016waw} and for $n=2$, they are the new integrable
vortex equations studied in this paper.
The question remains whether there are further integrable vortex
equations left in the Taubes-like class of vortex equations.

Abelian vortex equations on constant curvature Riemann surfaces, in
particular the ones studied in Ref.~\cite{Manton:2016waw}, have been
reinterpreted as flat non-Abelian Cartan
connections \cite{Ross:2021afj} by lifting the vortex equations to
three dimensional group manifolds.
It would be interesting to see whether the new integrable vortex
equations, found in this paper, have such interpretation and how their
differences manifest themselves in the Cartan connections.
For instance, how the difference in the conical deficits are realized
in the Cartan connections after lifting the equations.

Impurities and in particular magnetic impurities have been studied in
vortex systems and a nice interpretation of the impurity is given by a
set of coupled vortex equations, in the limit where one of the vortex
fields becomes much heavier than the other; the ``frozen'' vortex may
thus be interpreted in the remaining vortex equation as a magnetic
impurity \cite{Tong:2013iqa}.
The magnetic type of impurity has been generalized to all Manton's
five integrable vortex systems in Ref.~\cite{Gudnason:2021bkw}.
It would be interesting to study magnetic impurities in the newly
found vortex equations.

By allowing the metric to depend on the Higgs field, it is known that
other equations than the Liouville equation can be related to the
Taubes equation for Abelian Higgs vortices \cite{Dunajski:2012nv}.
It would be interesting to see what relations would be possible from
the vortex equations considered in this paper. 

Non-Abelian generalizations on the sphere for the known integrable
vortices have been considered in Ref.~\cite{Walton:2021dgi}.
It would be interesting to consider such generalizations for the new
vortex equations found in this paper.

\subsection*{Acknowledgements}
I thank Muneto Nitta for discussions during the initial stage of this
project and Nick Manton for comments on the manuscript.
S.~B.~G.~thanks the Outstanding Talent Program of Henan University and
the Ministry of Education of Henan Province for partial support.
The work of S.~B.~G.~is supported by the National Natural Science
Foundation of China (Grants No.~11675223 and No.~12071111).

\bibliographystyle{utphys}
\bibliography{refs}

\providecommand{\href}[2]{#2}\begingroup\raggedright\begin{thebibliography}{10}

\bibitem{Taubes:1979tm}
C.~H. Taubes, ``{Arbitrary N: Vortex Solutions to the First Order
  Landau-Ginzburg Equations},''
  \href{http://dx.doi.org/10.1007/BF01197552}{{\em Commun. Math. Phys.}
  {\bfseries 72} (1980) 277--292}.

\bibitem{JT}
A.~Jaffe and C.~Taubes, {\em Vortices and Monopoles -- Structure of Static
  Gauge Theories}.
\newblock Birkh{\"a}user, Boston, MA, USA, 1980.

\bibitem{MS}
N.~Manton and P.~M. Sutcliffe, {\em Topological Solitons}.
\newblock Cambridge University Press, Cambridge, UK, 2004.

\bibitem{YangsBook}
Y.~Yang, {\em Solitons in Field Theory and Nonlinear Analysis}.
\newblock 2001.

\bibitem{Witten:1976ck}
E.~Witten, ``{Some Exact Multi - Instanton Solutions of Classical Yang-Mills
  Theory},'' \href{http://dx.doi.org/10.1103/PhysRevLett.38.121}{{\em Phys.
  Rev. Lett.} {\bfseries 38} (1977) 121--124}.

\bibitem{Jackiw:1990tz}
R.~Jackiw and S.~Y. Pi, ``{Soliton Solutions to the Gauged Nonlinear
  Schrodinger Equation on the Plane},''
  \href{http://dx.doi.org/10.1103/PhysRevLett.64.2969}{{\em Phys. Rev. Lett.}
  {\bfseries 64} (1990) 2969--2972}.

\bibitem{Popov:2012av}
A.~D. Popov, ``{Integrable vortex-type equations on the two-sphere},''
  \href{http://dx.doi.org/10.1103/PhysRevD.86.105044}{{\em Phys. Rev. D}
  {\bfseries 86} (2012) 105044},
  \href{http://arxiv.org/abs/1208.3578}{{\ttfamily arXiv:1208.3578 [hep-th]}}.

\bibitem{Ambjorn:1988fx}
J.~Ambjorn and P.~Olesen, ``{Antiscreening of Large Magnetic Fields by Vector
  Bosons},'' \href{http://dx.doi.org/10.1016/0370-2693(88)90120-7}{{\em Phys.
  Lett. B} {\bfseries 214} (1988) 565--569}.

\bibitem{Manton:2016waw}
N.~S. Manton, ``{Five Vortex Equations},''
  \href{http://dx.doi.org/10.1088/1751-8121/aa5f19}{{\em J. Phys. A} {\bfseries
  50} no.~12, (2017) 125403}, \href{http://arxiv.org/abs/1612.06710}{{\ttfamily
  arXiv:1612.06710 [hep-th]}}.

\bibitem{Jackiw:1990aw}
R.~Jackiw and E.~J. Weinberg, ``{Selfdual Chern-Simons Vortices},''
  \href{http://dx.doi.org/10.1103/PhysRevLett.64.2234}{{\em Phys. Rev. Lett.}
  {\bfseries 64} (1990) 2234}.

\bibitem{Chen:2004xu}
H.-Y. Chen and N.~S. Manton, ``{The Kahler potential of Abelian Higgs
  vortices},'' \href{http://dx.doi.org/10.1063/1.1874334}{{\em J. Math. Phys.}
  {\bfseries 46} (2005) 052305},
  \href{http://arxiv.org/abs/hep-th/0407011}{{\ttfamily arXiv:hep-th/0407011}}.

\bibitem{Baptista:2012tx}
J.~M. Baptista, ``{Vortices as degenerate metrics},''
  \href{http://dx.doi.org/10.1007/s11005-014-0683-4}{{\em Lett. Math. Phys.}
  {\bfseries 104} (2014) 731--747},
  \href{http://arxiv.org/abs/1212.3561}{{\ttfamily arXiv:1212.3561 [hep-th]}}.

\bibitem{Contatto:2017alh}
F.~Contatto and M.~Dunajski, ``{Manton\textquoteright{}s five vortex equations
  from self-duality},'' \href{http://dx.doi.org/10.1088/1751-8121/aa8193}{{\em
  J. Phys. A} {\bfseries 50} no.~37, (2017) 375201},
  \href{http://arxiv.org/abs/1704.05875}{{\ttfamily arXiv:1704.05875
  [hep-th]}}.

\bibitem{Gudnason:2017jsn}
S.~B. Gudnason and M.~Nitta, ``{Some exact Bradlow vortex solutions},''
  \href{http://dx.doi.org/10.1007/JHEP05(2017)039}{{\em JHEP} {\bfseries 05}
  (2017) 039}, \href{http://arxiv.org/abs/1701.04356}{{\ttfamily
  arXiv:1701.04356 [hep-th]}}.

\bibitem{Bradlow:1990ir}
S.~B. Bradlow, ``{Vortices in holomorphic line bundles over closed Kahler
  manifolds},'' \href{http://dx.doi.org/10.1007/BF02097654}{{\em Commun. Math.
  Phys.} {\bfseries 135} (1990) 1--17}.

\bibitem{GH}
P.~Griffiths and J.~Harris, {\em Principles of Algebraic Geometry}.
\newblock John Wiley \& Sons, Inc., Hoboken, NJ, USA, 1994.

\bibitem{Kazdan:1974}
J.~Kazdan and F.~W. Warner, ``Curvature functions for compact 2-manifolds,''
  \href{http://dx.doi.org/10.2307/1971012}{{\em Annals of Mathematics}
  {\bfseries 99} (1974) 14}.

\bibitem{Ross:2021afj}
C.~Ross, ``{Cartan Connections and Integrable Vortex Equations},''
  \href{http://arxiv.org/abs/2112.08328}{{\ttfamily arXiv:2112.08328
  [math-ph]}}.

\bibitem{Tong:2013iqa}
D.~Tong and K.~Wong, ``{Vortices and Impurities},''
  \href{http://dx.doi.org/10.1007/JHEP01(2014)090}{{\em JHEP} {\bfseries 01}
  (2014) 090}, \href{http://arxiv.org/abs/1309.2644}{{\ttfamily arXiv:1309.2644
  [hep-th]}}.

\bibitem{Gudnason:2021bkw}
S.~B. Gudnason and C.~Ross, ``{Magnetic Impurities, Integrable Vortices and the
  Toda Equation},'' \href{http://dx.doi.org/10.1007/s11005-021-01444-8}{{\em
  Lett. Math. Phys.} {\bfseries 111} (2021) 100},
  \href{http://arxiv.org/abs/2105.01332}{{\ttfamily arXiv:2105.01332
  [math-ph]}}.

\bibitem{Dunajski:2012nv}
M.~Dunajski, ``{Abelian vortices from Sinh--Gordon and Tzitzeica equations},''
  \href{http://dx.doi.org/10.1016/j.physletb.2012.02.078}{{\em Phys. Lett. B}
  {\bfseries 710} (2012) 236--239},
  \href{http://arxiv.org/abs/1201.0105}{{\ttfamily arXiv:1201.0105 [hep-th]}}.

\bibitem{Walton:2021dgi}
E.~Walton, ``{Exotic vortices and twisted holomorphic maps},''
  \href{http://arxiv.org/abs/2108.00315}{{\ttfamily arXiv:2108.00315
  [math-ph]}}.

\end{thebibliography}\endgroup

\end{document}